# Modulation of switching dynamics in magnetic tunnel junctions for low-error-rate computational random-access memory

Yang Lv*, Brahmdutta Dixit*, and Jian-Ping Wang#
Electrical and Computer Engineering Department, University of Minnesota, Minneapolis, MN 55455, USA
*both authors contributed equally

**ABSTRACT**. The conventional computer architecture has been facing challenges answering the ever-increasing demands from emerging applications, such as AI, for energy-efficient computation and memory hardware systems. Computational Random Access Memory (CRAM) represents a true in-memory computing paradigm that integrates logic and memory functions within the same array. At its core, CRAM relies on Magnetic Tunnel Junctions (MTJs), which serve as the foundational building blocks for implementing both memory storage and logic operations. However, a key challenge in CRAM lies in the non-ideal error rates associated with switching dynamics of MTJs, necessitating innovative approaches to reduce errors and optimize logic margins. This work proposes a novel approach of utilizing the voltage-controlled magnetic anisotropy (VCMA) to steepen the switching probability transfer curve (SPTC), thereby significantly reducing the logic operation error rate in CRAM. Using several numerical modeling tools, we validate the effectiveness of VCMA in modulating the energy barrier and switching dynamics in MTJs. It is revealed that the VCMA effect significantly reduces the error rate of CRAM by 61.43% at a VCMA coefficient of 200 $fJ \cdot V^{-1}m^{-1}$ compared to CRAM without VCMA. The reduction of error rate is further rapidly amplified with an increasing TMR ratio. Furthermore, the introduction of the VCMA effect decreases the logic voltage ($V_{logic}$) required for logic operations in CRAM and results in reduction of energy consumption. Our work serves as a first exploration in reducing the error rate in CRAM by modifying SPTC in MTJs.

**Introduction:**

The ever-increasing demand for computational performance in domains such as artificial intelligence (AI), machine learning (ML), and data analytics has exposed critical inefficiencies in the traditional von Neumann computing architecture [1-5]. A major bottleneck arises from the physical separation of logic and memory units, which results in frequent and energy-consuming data transfers. This "memory bottleneck" is particularly problematic in data-intensive applications. To overcome this challenge, the concept of in-memory computing [6-14] has gained attraction. Among its variants, the computational random-access memory (CRAM) has emerged as a transformative paradigm [15-16].

CRAM can act as both a memory and a computing medium. Unlike conventional computer architecture, CRAM performs logic operations (computing) directly among, within, and by its cells. This eliminates the need for data leaving and entering the array, which constitutes a key feature as a true in-memory computing paradigm [17]. CRAM leverages magnetic tunnel junctions (MTJs), a mature spintronic device characterized by high endurance, non-volatility, and low energy consumption, to implement both memory storage and logic operations [18].

In addition to the performance advantages provided by emerging memory technologies, CRAM offers several fundamental benefits at both the circuit and architecture levels as shown in Figure 1: (1) it eliminates the significant energy and performance costs associated with data transfer between logic and memory; (2) it allows for random access of data for input and output operations; (3) it supports reconfigurability, enabling programmable logic operations such as AND, OR, NAND, NOR, and MAJ; and (4) it enables performance improvements through massive parallelism, as identical operations can be executed simultaneously across rows in the CRAM array when data is allocated efficiently [16].

Analysis and benchmarking suggest that CRAM holds great promise for achieving significant gains in logic operations and true-in memory operations. This is particularly beneficial for data-intensive, memory-centric, and power-sensitive applications, such as bioinformatics, image and signal processing, neural networks, and edge computing [19-21].

The CRAM array has many variants, including the 2T1M configuration, which is modified from the 1T1M structure of STT-MRAM to enable logic operations within the CRAM. During logic operations, a 'Y-shape' circuit is temporarily formed in the CRAM array, where the input MTJs contribute their states to control the current flow through the output MTJ. The CRAM logic operation, as illustrated in Figure 1, relies on the Voltage-Controlled Logic (VCL) as its key working principle. It leverages the thresholding effect observed during MTJ switching and the TMR effect of the MTJ. In its memory mode, the CRAM functions like a conventional memory, allowing data to be read or written via memory bit lines (MBL). In its logic mode, computations are performed using logic lines (LL) and bit select lines (BSL), where the resistance states of MTJs determine the current flow and influence logic outcomes.

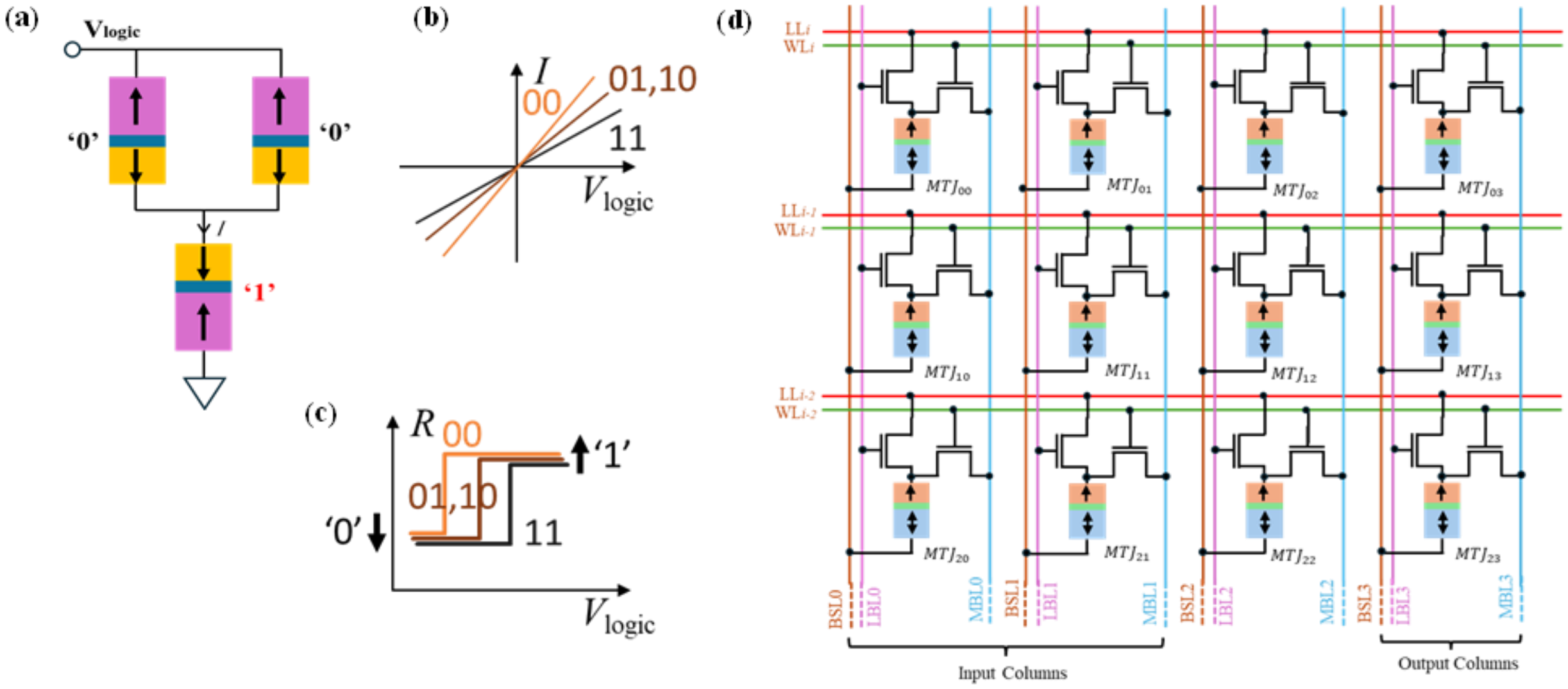

Figure 1. (a) Device-level CRAM illustration (P = '0' (free-layer arrow down, parallel to reference), AP = '1' (free-layer arrow up, antiparallel)) and corresponding (b) I-V and (c) R-V curve showcasing $V_{logic}$ for different logic operations and (d) STT-VCMA array.

Operations such as AND, OR, NAND, and NOR are executed by configuring the resistance states of MTJs in a row, with output state changes occurring when the resulting current exceeds a critical threshold. Inter-row logic operations are enabled by incorporating switches between rows, allowing data transfer and computation across rows, which is essential for multi-row operations like adders or multipliers. Throughout this work we map parallel alignment to logic '0' and antiparallel alignment to logic '1', as illustrated in Fig. 1. Additionally, CRAM supports high degrees of parallelism, allowing independent operations across multiple rows simultaneously. Efficient scheduling of logic operations ensures minimal data movement and increases computational throughput, making CRAM a robust platform for applications like neural inference engines and 2D convolution, while significantly reducing energy and delay overheads compared to traditional memory and processing architectures.

In order to reduce the likelihood of incorrect logic output due to various factors affecting the reliability of the operations, therefore CRAM error rate required to be minimized. This error rate exists primarily because CRAM logic relies on the precise manipulation of current and voltage levels to switch the states of MTJs. Basic contributors to errors include process variations, thermal noise, parasitic effects, and deviations in the TMR ratio, which can impact the critical switching currents. The effective ways to reduce the error rate include improving higher TMR ratios and lower resistance-area (RA) products. In this work we propose how VCMA effect can be implemented to improve error rate by modulating switching probability of MTJs. By dynamically modulating the energy barrier of MTJs, VCMA improves the steepness of the SPTC for an MTJ. First, we demonstrate how a high VCMA coefficient enhances SPTC, and then we incorporate it into CRAM to show how it improves the error rate and logic margin in NAND logic operations, enabling true in-memory computation. Therefore, the present work focuses on integrating VCMA-enhanced MTJs within CRAM arrays and examines their impact on switching dynamics, energy efficiency, and operational reliability.

**Methods:**

VCMA allows voltage modulation of PMA, specifically, when a positive voltage is applied to the MTJ, interfacial PMA is reduced [22]. This alteration occurs due to charge accumulation or depletion at the interface between the metal and barrier layer, which influences the spin-orbit coupling responsible for anisotropy changes. The relationship between the interfacial anisotropy energy density ($K_{int}$) and applied voltage (V) is expressed as:

$$K_{int} = K_{int,\ V=0} - \xi V / t_{ox} \qquad (1)$$

Here, $K_{int,\ V=0}$ denotes the anisotropy energy density in the absence of an applied voltage, $\xi$ represents the VCMA coefficient (a measure of sensitivity between magnetic anisotropy and the electric field), and $t_{ox}$ is the thickness of the spacer oxide layer between two ferromagnetic layers.

When the interfacial PMA is altered, it effectively modulates the energy barrier ($E_b$) of the free magnetic layer [23-24]. The thermal stability factor ($\Delta$) of an MTJ, a crucial parameter for data retention, can be expressed in terms of the energy barrier as:

$$\Delta = \frac{E_b}{k_B T} = \frac{(K_{int} - 2\pi M_s^2 t_F)A}{k_B T} \qquad (2)$$

Where $k_B$ is Boltzmann constant, $M_s$ is saturation magnetization, $t_F$ and A is thickness and cross-sectional area of free layer respectively and $T$ is the absolute temperature. The randomness in the magnetization dynamics, commonly referred to as thermal fluctuation, plays a significant role in influencing the switching behavior of MTJ. Thermal fluctuation can activate the magnetization's initial angle at the start of the writing process, potentially leading to either undesirable switching events or hindering the intended switching model the impact of thermal fluctuations [25].
Further, random thermal field is applied not only to the initial angle of magnetization but also to the effective anisotropy field at every time step during the HSPICE simulation.
This stochastic nature of the thermal field is best represented as a zero-mean Gaussian random distribution, with its standard deviation ($\sigma_{H_{th}}$) given by [26]:

$$\sigma_{H_{th}} = \sqrt{\frac{2 k_B \alpha T}{\mu_0 \gamma V_F M_s \delta t}} \qquad (3)$$

Here, $\alpha$ is the Gilbert damping coefficient, $\mu_0$ is the permeability of free space, $\gamma$ gyromagnetic ratio, $V_F$ denotes the volume of the free layer, $M_s$ represents the saturation magnetization, and $\delta t$ is the time step in the simulation. For the simulations conducted in this study, a value of $\sigma_{H_{th}} = 4.5mT$ was used to accurately capture the effects of thermal fluctuations under realistic conditions. This approach ensures the proper emulation of random thermal effects in Monte Carlo simulations.

**Magnetization Dynamics:**

The SPICE model utilized in this study is derived from the Landau–Lifshitz–Gilbert (LLG) equation, which encompasses the effects of precession, damping, and spin-transfer torque (STT). The equation is expressed as follows:

$$\left(1 + \frac{\alpha^2}{\gamma}\right)\frac{d\vec{M}}{dt} = -\vec{M} \times \vec{H}_{eff}(V) - \alpha\left(\vec{M} \times \left(\vec{M} \times \vec{H}_{eff}(V)\right)\right) + \frac{\hbar P J}{2 e t_F M_s} \times \vec{M} \times \left(\vec{M} \times \vec{M}_p\right) \qquad (4)$$

Here, $M$ represents the magnetization vector of the free layer, and $\vec{H}_{eff}(V)$ is the effective magnetic field, which depends on the applied voltage $V$. Other terms include $P$, the spin polarization factor, $J$, the current density, $t_F$ free layer thickness and $\vec{M}_p$ the fixed layer magnetization vector.

The effective magnetic field, $\vec{H}_{eff}(V)$, accounts for various components such as the external magnetic field ($\vec{H}_{ext}$), the demagnetization field ($\vec{H}_d$), thermal field ($\vec{H}_{th}$) and the voltage-controlled perpendicular anisotropy field $\vec{H}_{K\perp eff}(V)$ as the mathematical expression [27]:

$$\vec{H}_{eff}(V) = \vec{H}_{ext} + \vec{H}_d + \vec{H}_{th} + \vec{H}_{K\perp eff}(V) \qquad (5)$$

$$\vec{H}_{K\perp eff}(V) = \left(0\vec{x},\ 0\vec{y},\ \left(\frac{2K_i(V)}{\mu_0 M_s t_F}\right) m_z \vec{z}\right) \qquad (6)$$

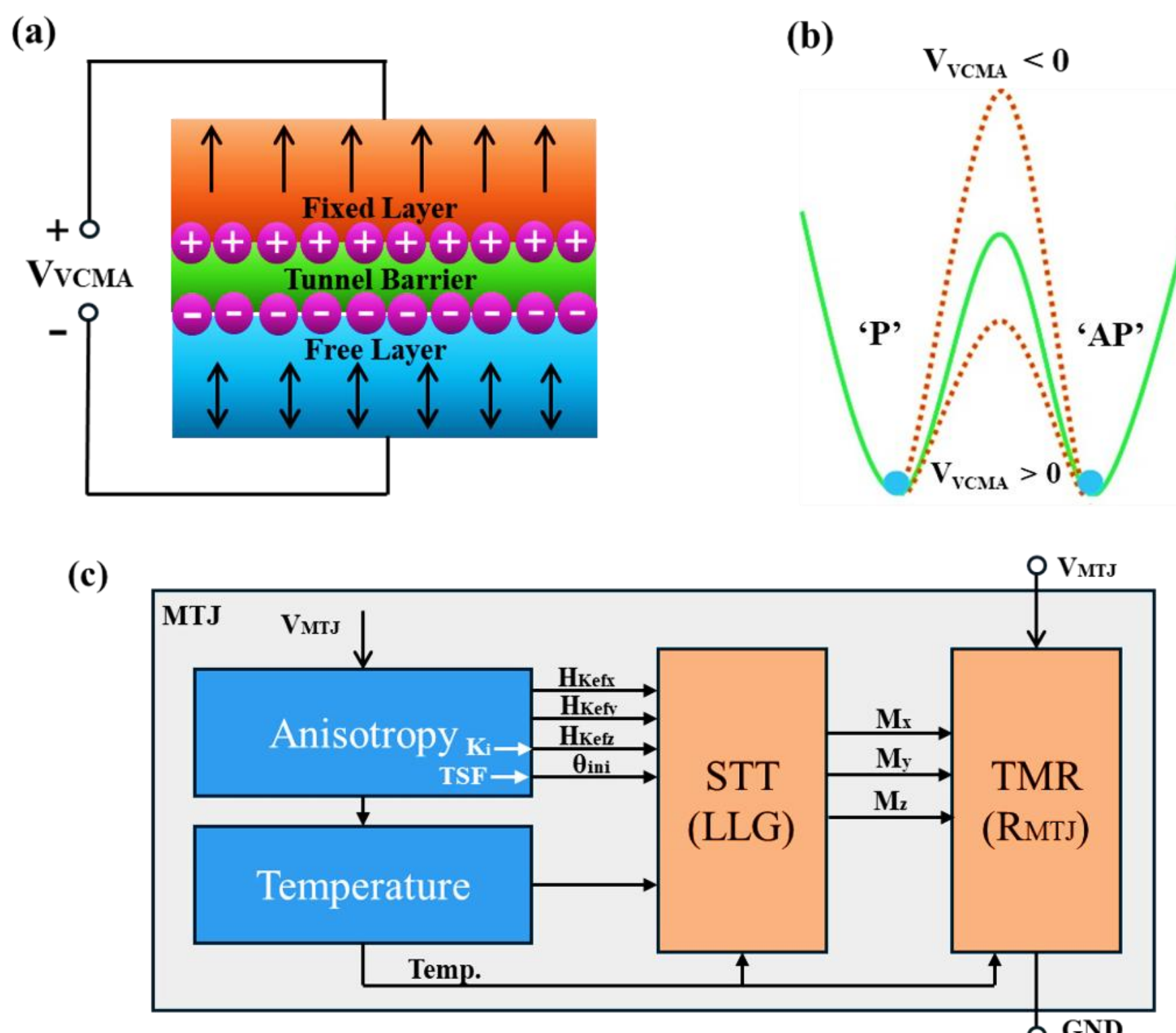

Figure 2: (a) Device structure of STT+VCMA MTJ . (b) The energy barrier of a VCMA-based MTJ is influenced by the applied voltage ($V_{VCMA}$). A reduction in the energy barrier enables the magnetization to oscillate between its parallel and antiparallel configurations. (c) VCMA-STT MTJ SPICE compact model capturing the interplay of anisotropy, temperature effects, STT dynamics, and TMR for simulating VCMA-MTJ behavior.

Where, $\mu_0$ is the permeability of free space. Further, the VCMA effect influences the thermal stability factor (TSF), which in turn impacts the magnetization dynamics described in the LLG equation.

The initial magnetization angle (θ) is treated as a stochastic variable and is modeled using the Fokker–Planck distribution [28]:

$$PDF(\theta)|t = 0 = \frac{\exp\left(-TSF \cdot sin^2\theta\right)}{\int_0^{\pi} sin\theta \cdot exp(-TSF \cdot sin^2\theta) d\theta} \qquad (7)$$

Here, denotes the thermal stability factor, which characterizes the retention ability of the free layer magnetization. The temperature dependence and tunneling magnetoresistance (TMR) of the MTJ are modeled using modified Julliere's formula [29]:

$$TMR(T,V) = \frac{2P^2(1-\alpha_{sp}\,T^3/2)^2}{1-P_0^2(1-\alpha_{sp}\,T^3/2)^2} \cdot \frac{1}{1+(V/V_0)^2} \quad (8)$$

Where, $P_0$ is the polarization factor, $\alpha_{sp}$ is a temperature-dependent constant, and $V_0$ is the bias voltage at which TMR is halved. The VCMA-based MTJ SPICE package as illustrated in Fig. 2(c) integrates these equations with four sub-circuits: anisotropy, LLG dynamics, TMR, and temperature effects, enabling realistic simulation of VCMA-MTJ behavior [30-31].

**CRAM Logic Operation Error Rate:**

To analyze CRAM logic operation error rates, we employ a modeling approach like that used in previous studies on MTJ-based CRAM hardware [15-16]. As CRAM logic operation follows the VCL principle, where MTJs function as both memory storage and logic components. During a logic operation, selected MTJs in a row are temporarily connected via a logic line (LL), and the switching behavior of the output MTJ is determined by the resistance states of input MTJs and the applied voltage.

The error rate in CRAM logic operations arises from the probabilistic nature of MTJ switching, which has dependency on thermal fluctuations, process variations, and deviations in the TMR ratio. The switching probability of the output MTJ is not a step function but follows a sigmoidal SPTC. The steepness of this curve is a critical factor in determining the error rate.

In this work, we have multiple numerical simulations to obtain SPTCs and assess error rates. The methodology consists of three key steps:

1. Modeling the voltage-dependent tunneling TMR effect on MTJ.
2. Solving the voltage distribution across LL during operations.
3. Simulating the magnetization switching probability of the output MTJ, incorporating thermal noise and variations to evaluate logic accuracy.

The error rate is calculated by comparing the simulated statistical expectations of the output state, ⟨Dout⟩, with the expected Boolean logic truth table. The error rate metric quantifies the likelihood of an incorrect logic output, which is minimized by optimizing TMR ratios, voltage parameters ($V_{logic}$), and switching dynamics.

This methodology provides a realistic and experiment-based approach to assessing and improving CRAM manufacturability, ensuring its viability for energy-efficient, high-performance computing applications.

Due to the need to capture extremely rare switching events required to estimate switching probabilities on the order of $10^{-6}$, and the constraints of limited computational resources, we adopt a macrospin model and assume a rectangular MTJ shape for simplicity. This simplification allows for modeling the statistical initial angel distribution and real-time thermal noise in magnetic dynamics. However, for the specific dimensions considered in our work, the model is expected to capture the qualitative features of the SPTCs accurately, with only minor quantitative deviations compared to a micromagnetic model [32-34], which includes edge effects of the MTJ.

The device parameters adopted are listed in TABLE 2.1. Note that we adopted VCMA coefficient, ξ, of 200 $fJ{\cdot}V^{-1}{\cdot}m^{-1}$ to represent a reasonably optimistic outlook, considering experimental observations [35-41] range from 100 up to 370 $fJ{\cdot}V^{-1}{\cdot}m^{-1}$ and theoretical predictions pointing to 1000 $fJ{\cdot}V^{-1}{\cdot}m^{-1}$ in highly relevant materials and devices.

**Results and Discussion:**

This section presents the results of our study and a discussion of the observed trends and their implications.

**Switching Probability Transfer Curves with Various Thermal Stability Factors:**

The SPTC for different Δ were analyzed under normalized voltage conditions. Figure 3 shows that higher thermal stability factors modulate the steepness of SPTCs. Specifically, as Δ increases, the switching threshold becomes more well-defined, reducing the overlap region and thereby minimizing switching errors by much more deterministic switching. This behavior confirms that increasing thermal stability is critical for achieving higher steepness in switching probability. For Figure 3 we employ write pulse width of $t_p$ = 1 ms so that the curves show more obvious steepness trends. However, for the rest of the results, the pulse width is 1 ns. The plots in figure 3 are meant solely to qualitatively illustrate how the slope of the switching-probability curve changes with the thermal-stability factor Δ. Each curve is plotted against the normalized voltage $V_{MTJ}/V_{50}$ (Δ), where $V_{50}$ is the voltage at which the switching probability equals 0.5 for the given Δ, so that the influence of Δ on the curve steepness can be compared independently of its absolute voltage offset.

**TABLE 2.1 Device Parameters**

| Device Parameter | Description | Default value |
|---|---|---|
| $\mathbf{L_x}$ | Free Layer Width (nm) | 45 |
| $\mathbf{L_y}$ | Free Layer Length (nm) | 45 |
| $\mathbf{t_F}$ | Free Layer Thickness (nm) | 0.75 |
| $\mathbf{t_{ox}}$ | Oxide Layer Thickness | 1 |
| $\mathbf{t_C}$ | Critical Layer Thickness | 1.5 |
| $\mathbf{RA_P}$ | Resistance-Area-Product ($\Omega - \mu m^2$) | 5 |
| $\mathbf{M_{S0}}$ | Saturation Magnetization (KA/cm) at 0K | 950 |
| $\mathbf{P_0}$ | Polarization Factor at 0K | 0.54 |
| $\boldsymbol{\alpha}$ | Damping Factor | 0.02 |
| $\boldsymbol{\xi}$ | VCMA Coefficient ($fJ \cdot V^{-1} \cdot m^{-1}$) | 0, 200 |
| **TSF** | Thermal Stability Factor | 45.7 |
| $\mathbf{H_{EXT}}$ | External Magnetic Field (mT) | 0 |
| **T** | Temperature (K) | 300 |

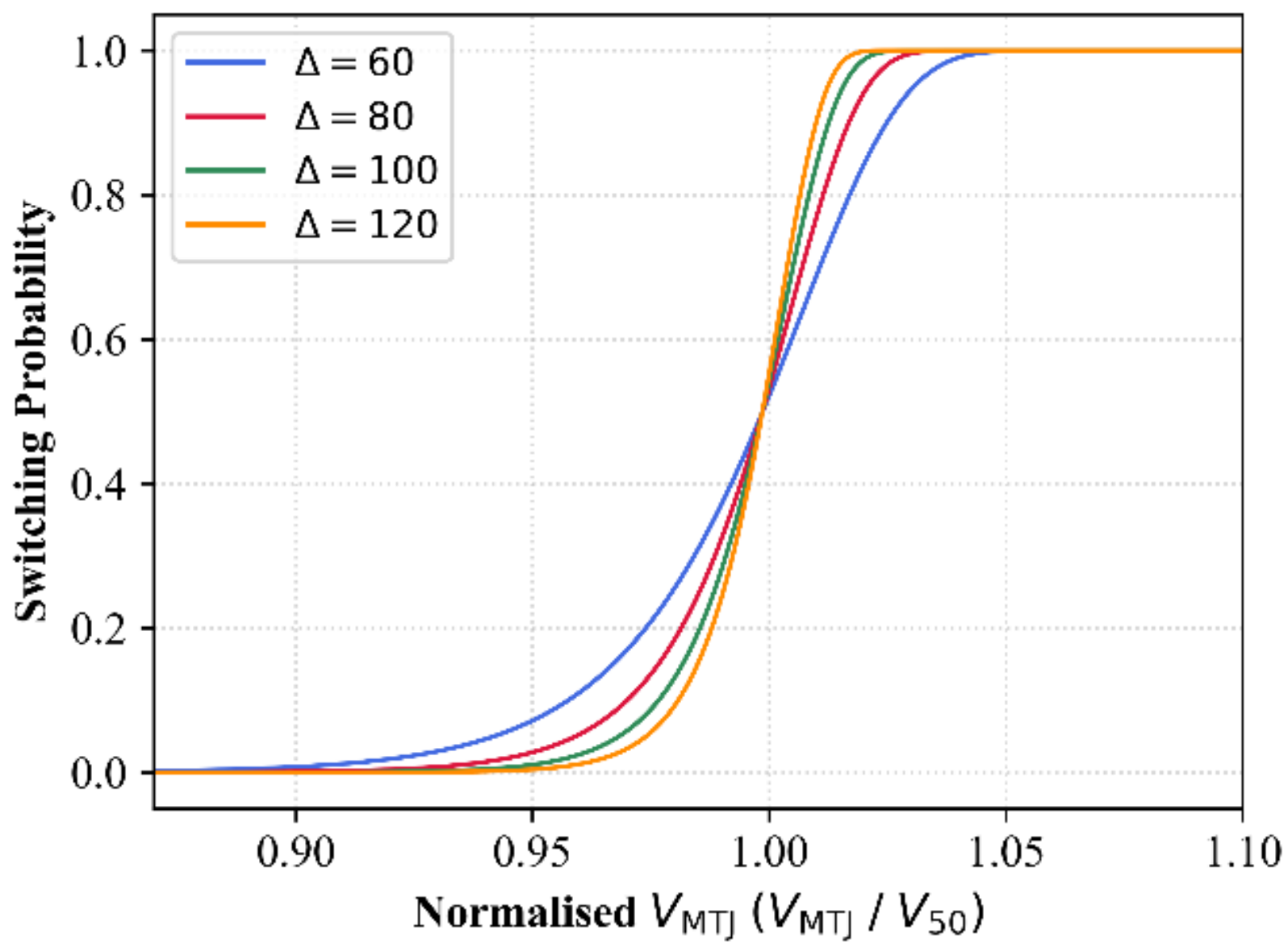

Figure 3: Qualitative illustration of thermal stability factors modulating the steepness of SPTCs. The pulse width is 1 ms.

**Magnetization Switching Dynamics:**

The magnetization dynamics under a 1 ns voltage pulse were analyzed as shown in Figure 4. The trajectories of $M_x$, $M_y$, and $M_z$ reveal precessional switching behavior, with the free layer magnetization transitioning between states. The oscillatory nature of $M_x$ and $M_y$ reflects the damping effects, while $M_z$ demonstrates the successful stabilization of the magnetization switching in the desired state. Unless otherwise noted, all subsequent simulations including the trajectories shown in Figure 4 use a 1ns write pulse. The device parameters adopted are listed in TABLE 1. Note that we adopted VCMA coefficient, ξ, of 200 $fJ \cdot V^{-1} \cdot m^{-1}$ to represent a reasonably optimistic outlook, considering experimental observations range from 100 up to 370 $fJ \cdot V^{-1} \cdot m^{-1}$ and theoretical predictions pointing to 1000 $fJ \cdot V^{-1} \cdot m^{-1}$ in highly relevant materials and devices.

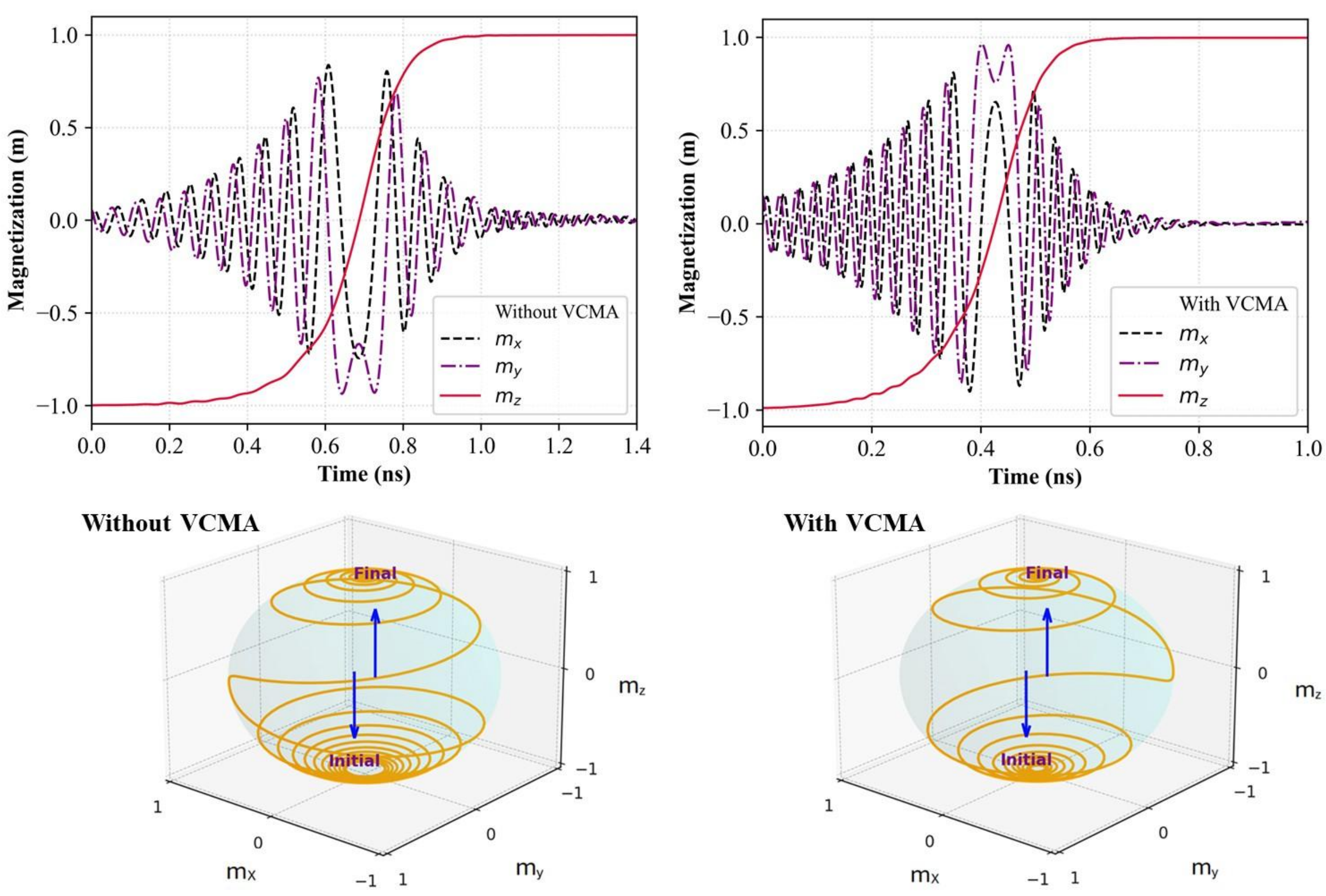

Figure 4: Simulation result of magnetization switching dynamics ($V_{STT}$ = 1.25V (Without VCMA) and $V_{STT}$ = 0.95V (With VCMA) with pulse width of 1ns.).

**SPTCs with and without VCMA Effect:**

The impact of VCMA on SPTCs was evaluated as shown in Figure 5. Without VCMA, the switching probability curve is less steep, indicating a broader transition region and higher error susceptibility. In contrast, the inclusion of VCMA results in a steeper SPTC, which could lead to a reduction in error rate in CRAM, as reasoned previously. This is attributed to the modulation of the energy barrier by VCMA, which alters the dynamics and switching trajectory of magnetization.

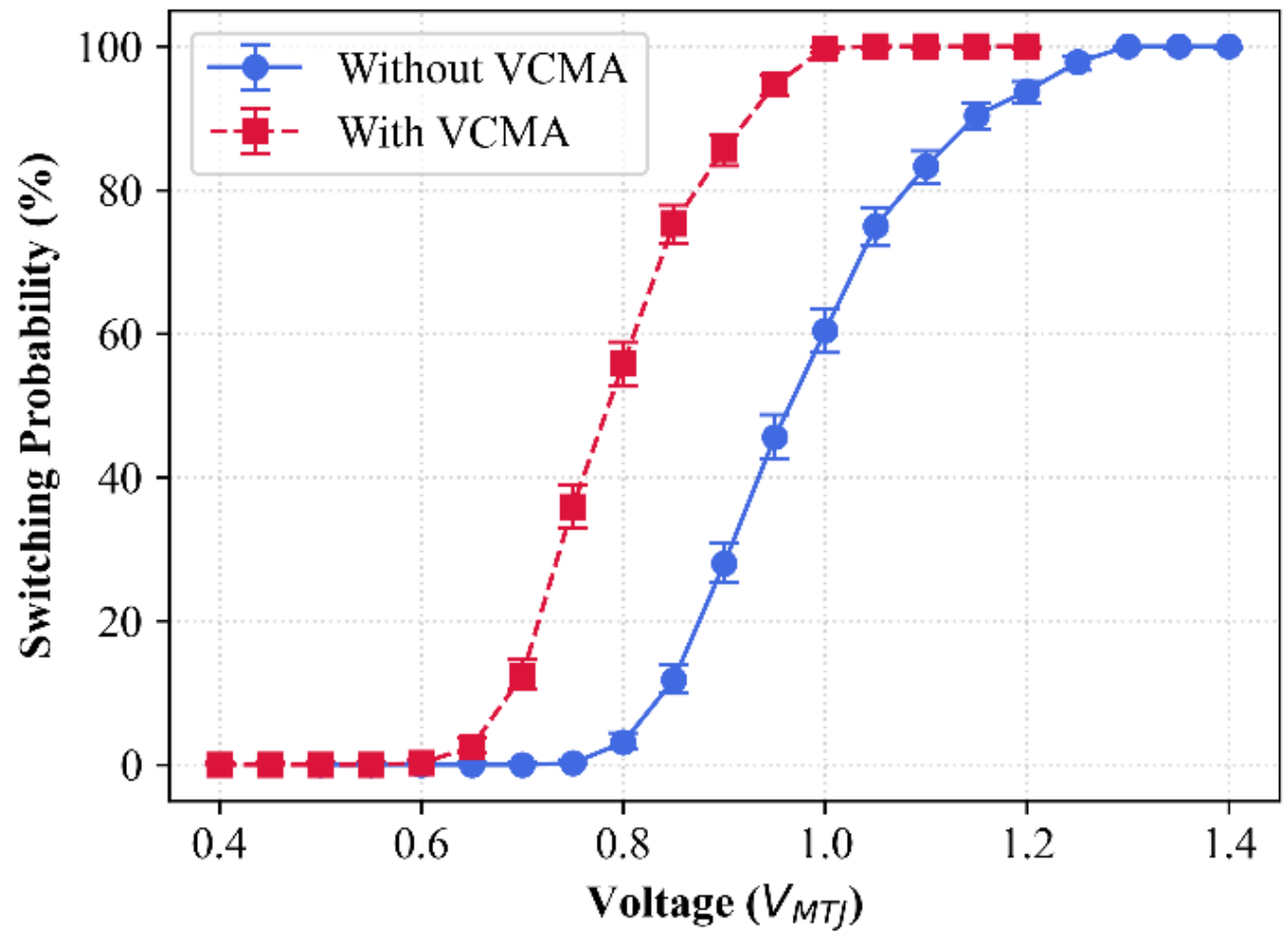

Figure 5: Comparison of SPTC plots: with VCMA ("VCMA = 200") and without VCMA ("VCMA = 0") at pulse width of 1ns (Each data point corresponds to Monte-Carlo simulations for 1000 switching trials).

**Probability and Complement Analysis:**

Steeper STPCs are crucial for reducing CRAM error rates. Being steeper is manifested as the switching probability stays more closely to 0 at low voltage and approaches 1 more quickly at high voltage. The results shown in Figure 6 stem from the same data shown in Figure 5, but are purposefully transformed and presented to show that SPTCs with VCMA achieve faster convergence to switching probabilities of 0 and 1 compared to those without VCMA.

The horizontal coordinates of the data points are normalized by the voltage needed to achieve 50% switching probability. The vertical coordinates are in logarithmic scale to further emphasize the effect of interest. The 'original' curves are plotted to mainly show how well the switching probability stays near 0 at low voltages. While the 'complementary' curves (1 - SPTC) exhibit how quickly the switching probability approaches to 1 at higher voltages. It is clear that the pair ('original' and 'complementary') for the case with VCMA show a narrow bell shape, indicating that the STPC is much steeper and adheres to 0 and 1 more aggressively.

**CRAM Output State:**

Dout is the Boolean state of the output MTJ, representing the result of a CRAM logic operation, such as NAND. The logic operation outcome is probabilistic due to the stochastic switching behavior of MTJs. Therefore, ⟨Dout⟩, the statistical average of Dout, is introduced to represent the average or statistical expectation of the output of a CRAM logic operation. The probabilistic behavior of an MTJ is captured through SPTC, allowing the calculation of ⟨Dout⟩ as a function of Vlogic for a CRAM logic operation. It is achieved by our circuit model, which also include the model of TMR and takes the SPTC as input. In other words, the SPTC is transformed into input-state-dependent ⟨$D_{out}$⟩ as a function of $V_{logic}$ by the circuit model. This relationship is critical because deviations of ⟨Dout⟩ from the ideal Boolean logic state (0 or 1) directly determine the CRAM logic operation error rate. The error rate is a function of $V_{logic}$, and it is determined by finding the minimum value among those with input '00', '01', and '10', and the value with input '11' subtracted from 1.0 at each $V_{logic}$ value for NAND operations. The best achievable error rate at a specific $V_{logic}$ values can then be identified. And for simplicity we refer to this minimum error rate interchangeably as error rate. Such dependencies with and without VCMA are illustrated in Figure 7.

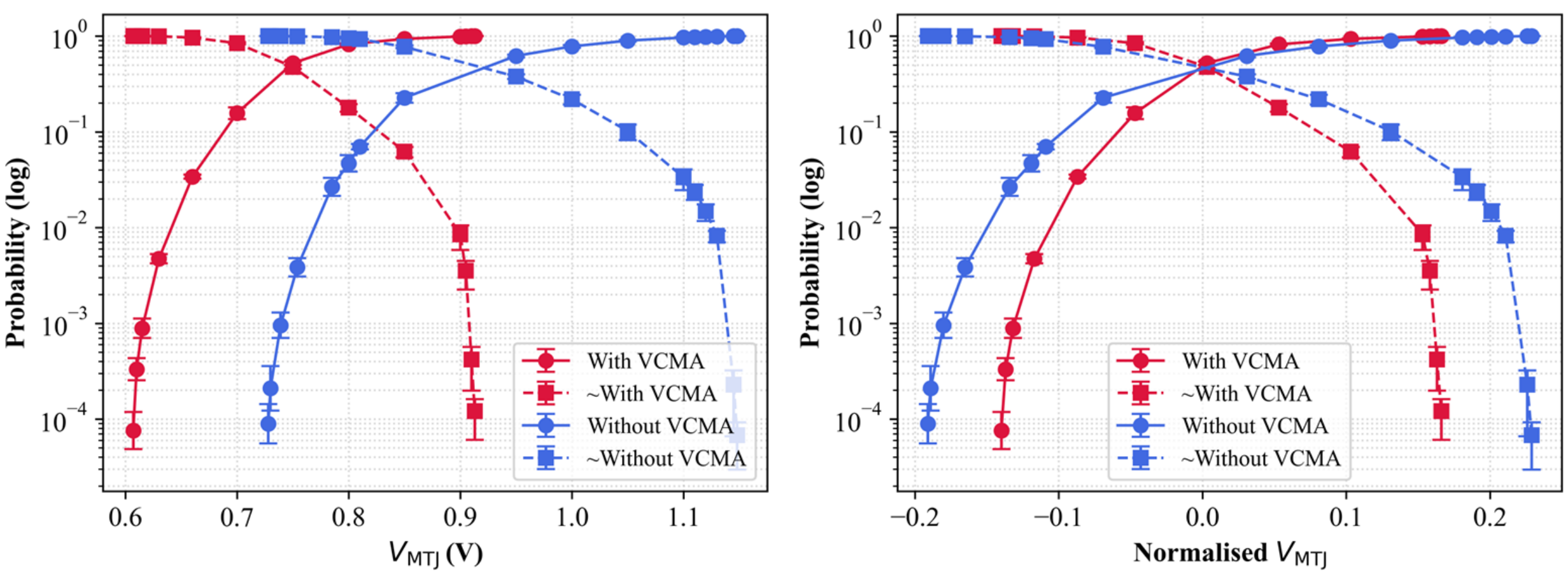

Figure 6: SPTCs with and without VCMA Effect

The statistical average outputs of a NAND logic operation of CRAM was analyzed with and without VCMA as illustrated in Figure 7. The VCMA-enhanced configuration required a lower logic voltage ($V_{logic}$) to achieve optimal error rates. Specifically, the logic voltage decreased from 1.801 V (without VCMA) to 1.458 V (with VCMA), accompanied by a reduction in the error rate from 26.33% to 17.25%. This demonstrates the dual benefit of VCMA in improving logic reliability and reducing energy consumption.

**CRAM Error Rate vs. TMR Ratio:**

The relationship between CRAM error rate and Tunnel Magnetoresistance (TMR) ratio was investigated (Figure 8). At a fixed TMR ratio, the error rate is consistently lower in VCMA-enhanced MTJs. For a TMR ratio of 200%, the error rate without VCMA was observed at $1.03\times10^{-1}$, while with VCMA, it was significantly reduced to $3.98\times10^{-2}$, amounting to a substantial decrease of 61.4%.

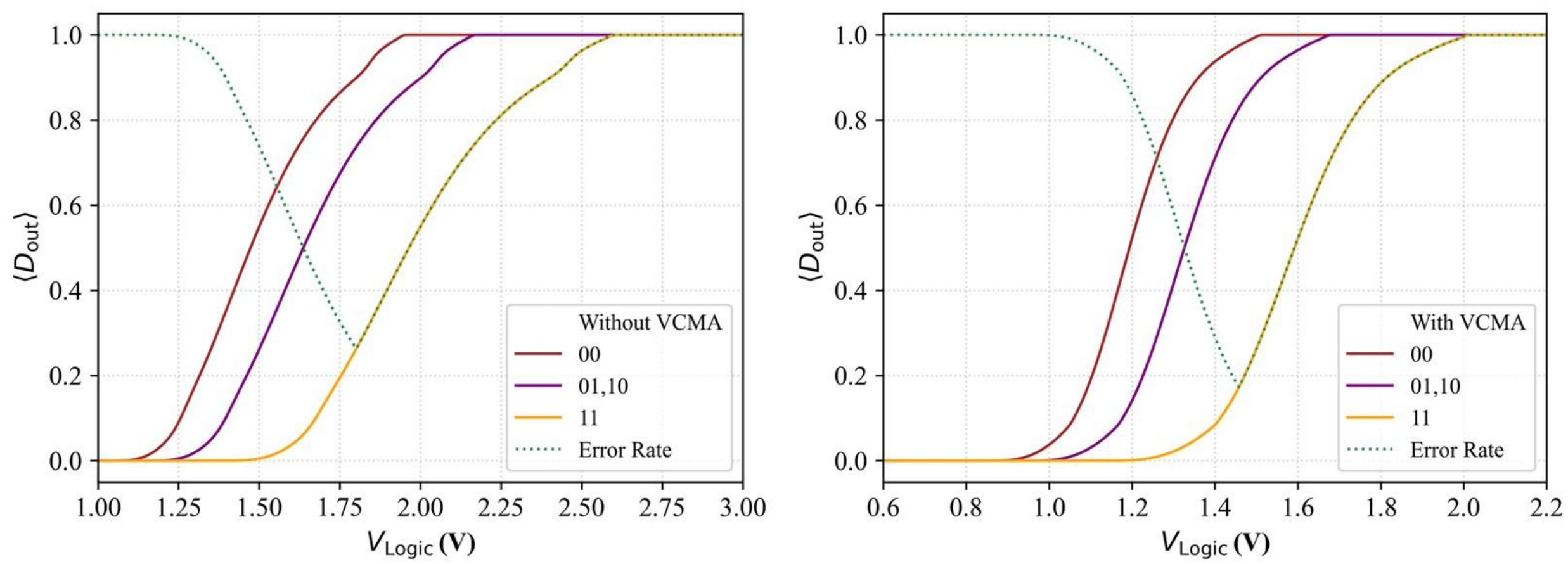

Figure 7: Illustration of Logic Voltage ($V_{logic}$) vs. output ($D_{out}$) with and without VCMA

Additionally, to achieve the same error rate with VCMA at 200% TMR ratio, a significantly higher (approximately 330%) TMR ratio is required without VCMA. For TMR ratios exceeding 200%, the error rate improvement becomes even more pronounced. This confirms that VCMA provides complementary advantages to existing techniques for error reduction. And employing our proposed technique can provide margins for error rate requirements, which could be traded off for requirements on TMR ratios, or other metrics.

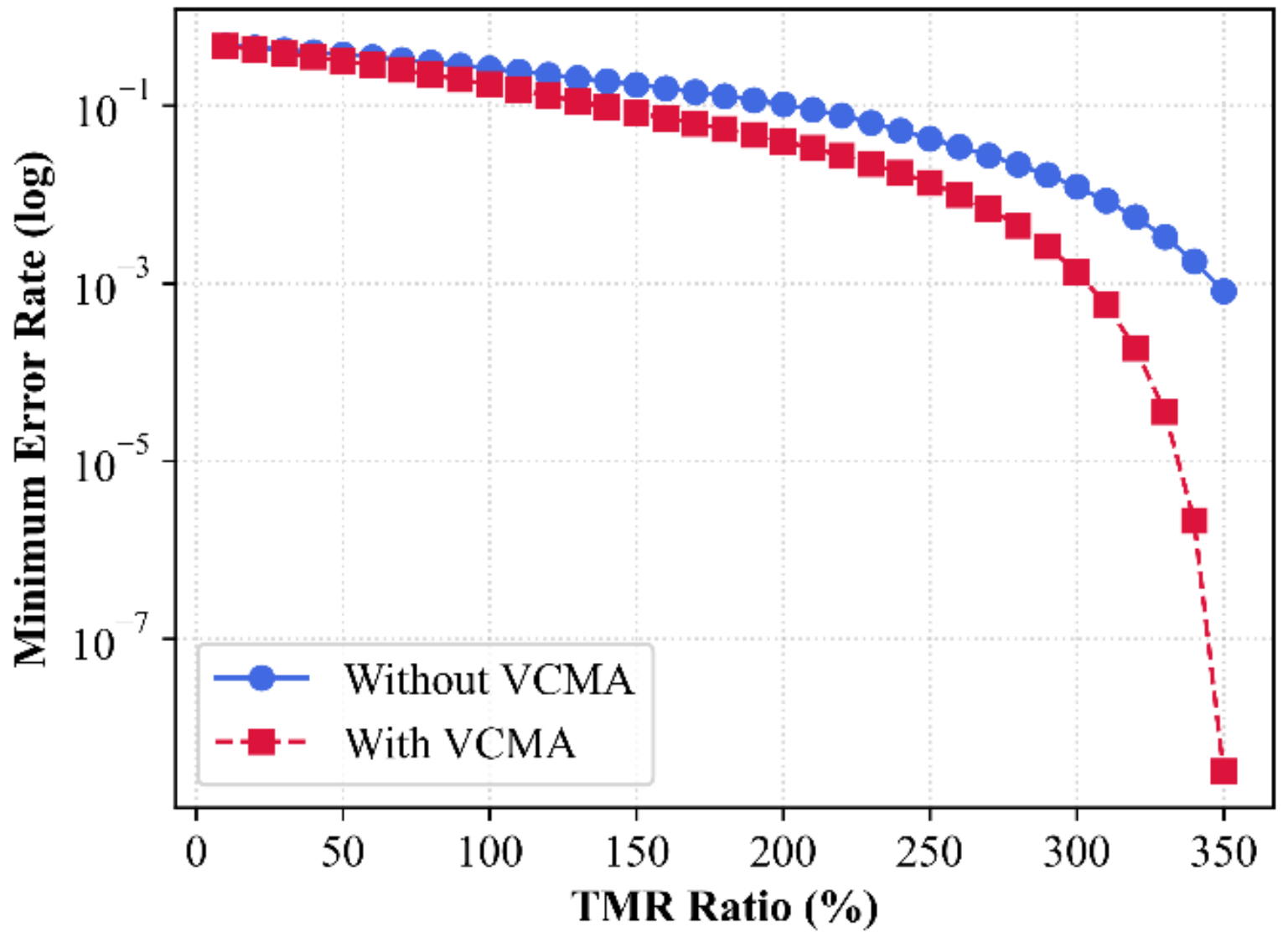

Figure 8: CRAM error rate vs. TMR ratio, with and without VCMA

**CRAM Logic Operation Energy Consumption:**

The energy consumption of CRAM logic operations was evaluated as demonstrated in Figure 9. At a TMR ratio of 200%, VCMA reduces energy consumption from $11{\times}10^{-13}$ J (without VCMA) to $7{\times}10^{-13}$ J, representing a significant improvement i.e. VCMA effect decreases the energy consumption of logic operations by approximately $4{\times}10^{-13}$ J. This energy reduction is directly linked to the decreased logic voltage requirement and highlights the efficiency gains achievable with VCMA. It is worth noting that results shown in this work are intended to demonstrate the validity and effectiveness of the proposed technique. Although we adopt many parameters based on experimental results, they are not necessarily optimized or optimistic for any performance metric, such as error rate, TMR ratio, energy efficiency.

**Discussions and Conclusion:**

This work presents a novel approach that leverages the VCMA effect to address a critical engineering bottleneck preventing the practical deployment of CRAM technology: reducing logic operation error rates. The results demonstrate significant benefits of integrating VCMA into MTJs for CRAM applications. By dynamically modulating the energy barrier, VCMA improves switching dynamics, reduces error rates, and lowers energy consumption. Key findings include sharper SPTC, faster convergence to switching states, and reduced logic voltage requirements, enabling enhanced reliability and efficiency. This work provides direct implications for advancing emerging technologies such as energy-efficient spintronic devices and AI hardware, highlighting its significance from both application and system integration perspectives. Note that due to the symmetry of VCMA's voltage dependence, our approach of improving error rate is limited to a certain subgroup of logic operations offered by CRAM. However, by achieving the opposite sign of VCMA coefficient, it is possible to target a specific logic operation subgroup.

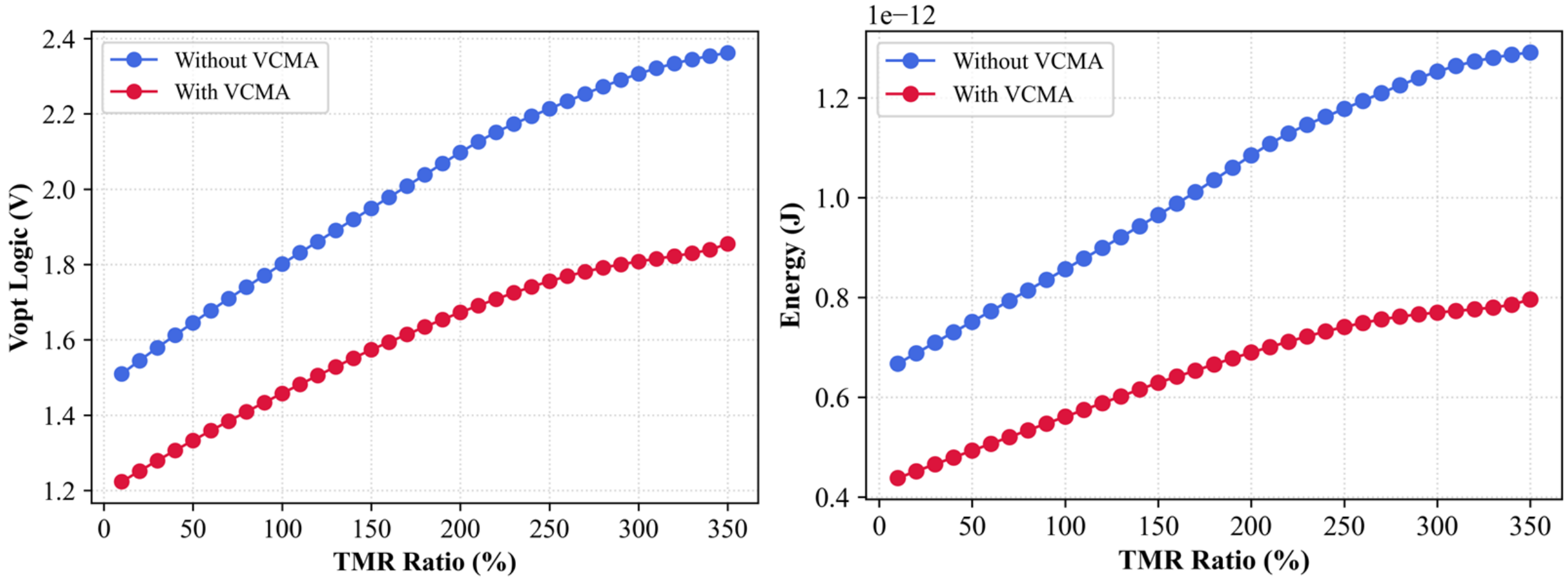

Figure 9: Illustration of $V_{logic}$ and Energy consumption variation with effective TMR ratio for without VCMA and with VCMA CRAM

At a TMR ratio of 200%, VCMA reduced the required logic voltage by approximately 20% and lowered energy consumption by $4\times10^{-13}$ J, while decreasing the error rate from 26.33% to 17.25%. These improvements highlight VCMA's potential to address critical challenges in high-performance, energy-efficient true in-memory systems.

This work underscores the importance of VCMA in advancing CRAM technology and provides a foundation for further research on leveraging spintronic devices for scalable and low-power computing applications. In conclusion, recent advancements in VCMA and voltage-controlled exchange coupling [42] (VCEC) present promising avenues to further reduce error rates and energy consumption in CRAM. Although CRAM implementations leveraging VCEC have yet to be demonstrated experimentally, the potential is substantial, given VCEC's demonstrated capability for ultrafast magnetization reversal within sub-100 picoseconds and energy-efficient stochastic switching suitable for neuromorphic computing applications [43,44].

Additionally, significant enhancements in VCMA efficiency achieved through electron depletion at the CoFeB/MgO interface via high work-function alloy underlayers have markedly increased VCMA coefficients [45]. Furthermore, novel composite heavy-metal tri-layer structures exploiting electron depletion, opposite spin Hall angles and unconventional z-spin enable field-free magnetization switching, aligning well with existing MRAM fabrication processes and further underscoring the feasibility of integrating advanced VCMA and potential VCEC methods to optimize CRAM performance and reliability [46-48].